\documentclass{aa}  

\usepackage{graphicx}

\usepackage{txfonts}
\newcommand{\urm}{$\rm rad\,m^{-2}$}
\newcommand{\lss}{{\it XMM}-LSS}
\usepackage[colorlinks,linkcolor=blue,anchorcolor=blue,citecolor=blue]{hyperref}
\usepackage{cleveref}
\begin{document}

   \title{Probing magnetic fields in the circumgalactic medium using polarization data from MIGHTEE}


    \author{K. Böckmann\inst{1}
    \and M. Brüggen\inst{1}
    \and V. Heesen\inst{1}
    \and A. Basu\inst{2}
    \and S. P. O'Sullivan\inst{3}
    \and I. Heywood\inst{4,5,6}
    \and M. Jarvis \inst{7,8}
    \and A. Scaife\inst{9,10}
    \and J. Stil\inst{11}
    \and R. Taylor\inst{12}
    \and N. J. Adams\inst{9}
    \and R. A. A. Bowler\inst{9}
    \and M. N. Tudorache \inst{7}
    }

   \institute{Hamburger Sternwarte, Universität Hamburg, Gojenbergsweg 112, 21029 Hamburg, Germany
   \and 
   Thüringer Landessternwarte, Sternwarte 5, 07778 Tautenburg, Germany
   \and
   Departamento de Física de la Tierra y Astrofísica \& IPARCOS-UCM, Universidad Complutense de Madrid, 28040 Madrid, Spain
   \and
   Astrophysics, Department of Physics, University of Oxford, Keble Road, Oxford, OX1 3RH, UK
   \and
   Department of Physics and Electronics, Rhodes University, PO Box 94, Makhanda, 6140, South Africa
   \and
   South African Radio Astronomy Observatory, 2 Fir Street, Observatory 7925, South Africa 
    \and
    Astrophysics, Department of Physics, University of Oxford, Keble Road, Oxford, OX1 3RH, UK
    \and
    National Radio Astronomy Observatory, 1003 Lopezville Road, Socorro, NM 87801, USA
   \and
    Jodrell Bank Centre for Astrophysics, Department of Physics and Astronomy, School of Natural Sciences, The University of Manchester, Manchester, M13 9PL, UK
   \and
    The Alan Turing Institute, Euston Road, London, NW1 2DB, UK
   \and 
   Department of Physics and Astronomy, The University of Calgary, 2500 University Drive NW, Calgary AB T2N 1N4, Canada
   \and
   Inter-Unversity Institute for Data Intensive Astronomy, Cape Town, South Africa
   }

   \date{08/22/2023}

 
  \abstract
   {The properties and evolution of magnetic fields surrounding galaxies are observationally largely unconstrained. The detection and study of these magnetic fields is important to understand galaxy evolution since magnetic fields are tracers for dynamical processes in the circumgalactic medium (CGM) and can have a significant impact on the evolution of the CGM.}
   {The Faraday rotation measure (RM) of the polarized light of background radio sources passing through the magnetized CGM of intervening galaxies can be used as a tracer for the strength and extent of magnetic fields around galaxies.}
   {We use rotation measures observed by the MIGHTEE-POL (MeerKAT International GHz Tiered Extragalactic Exploration POLarisation) survey by MeerKAT in the \lss\ and COSMOS fields to investigate the RM around foreground star-forming galaxies. We use spectroscopic catalogs of star-forming and blue cloud galaxies to measure the RM of MIGHTEE-POL sources as a function of the impact parameter from the intervening galaxy.  In addition, we examine the dependence of the RM on redshift. We then repeat this procedure using a deeper galaxy catalog with photometric redshifts.}
   {For the spectroscopic star-forming sample we find a redshift-corrected |RM| excess of $5.6\pm 2.3$\,\urm\ which corresponds to a $2.5\sigma$ significance around galaxies with a median redshift of $z = 0.46$ for impact parameters below 130\,kpc only selecting the intervenor with the smallest impact parameter. Making use of a photometric galaxy catalog and taking into account all intervenors with $M_g < -13.6$ mag, the signal disappears.
   We find no indication for a correlation between redshift and RM, nor do we find a connection between the total number of intervenors to the total |RM|.}  
   {We have presented tentative evidence that the CGM of star-forming galaxies is permeated by coherent magnetic fields within the virial radius. We conclude that mostly bright, star-forming galaxies with impact parameters less than 130 kpc significantly contribute to the RM of the background radio source.} 

   \keywords{cosmic rays -- galaxies: magnetic fields -- galaxies: fundamental parameters -- galaxies: ISM -- radio continuum: galaxies}

   \maketitle
%
\section{Introduction}

The diffuse gas embedded between the interstellar medium (ISM) and the baryon-rich intergalactic medium (IGM) of a galaxy is known as the circumgalactic medium (CGM) which typically extends up to the virial radius of galaxies $\approx$200 kpc. The existence of this extended gaseous halo is a fundamental prediction of galaxy formation theory \citep[e.g.][]{Tumlinson_2017}. Observations and simulations from across the whole electromagnetic spectrum and redshifts suggest that the CGM gas has a major impact on galaxy evolution and in the chemical history and evolution of a galaxy \citep[e.g.][]{2022PhR...973....1D}. Firstly, it receives enriched material that was expelled in the form of outflows; secondly, it also acts as a reservoir of fuel for future star formation, including the infalling IGM gas \citep{Machado_2018}. One powerful tool that is used to study the tenuous multi-phase CGM are transverse absorption-line studies
using background quasars,
which enables one to determine the relative abundance of various elements and thus the metallicity of the gas \citep{Mintz2020, DeFelippis2021}.

Magnetic fields are an important non-thermal component in and around galaxies that affect the dynamics and structure of the CGM. In particular, they are believed to play an important role in transport of materials to and from disks into the CGM \citep{Aramburo-Garcia_2023} and affect how gas is accreted onto galaxies \citep{Heesen_2023}. However, little is known about the dynamical importance of the magnetic fields and their evolution in galaxies and the surrounding CGM over cosmic time. Moreover, it is not yet understood how the CGM gets magnetized in the first place. Two possible scenarios have been discussed: (i) magnetic fields could be generated by small-scale dynamo effects \citep{Pakmor_2017, Pakmor_2020} and (ii) magnetic fields could be generated in the disk and subsequently transported out via galactic winds and outflows \citep{Peroux_2020}. In the former case, we expect the resulting magnetic fields to be turbulent with no ordered fields on the size of the CGM. However, the turbulent magnetic fields can be converted to anisotropic fields by shear flows \citep{Fletcher_2011}. If galactic winds and feedback are the origin of the magnetized CGM, a strong azimuthal dependence with the galactic disk is expected because the outflows are usually along the minor axis. This is valid for both, stellar and AGN (Active Galatic Nuclei) driven feedback \citep{Thomas_2022, Pillepich_2021}. 

Magnetic fields in nearby (distance < 50 Mpc) star-forming disc galaxies have been studied in some detail, with most studies indicating the existence of a galactic dynamo \citep{Beck_1996, Beck_2015}. For more distant galaxies at redshift $z>0.1$, the data is scarcer with only a few studies existing \citep{Oren_1995, Bernet_2008, Farnes_2014, Kim_2016, Mao_2017, Lan_2020}. Measuring the magnetic fields directly via synchrotron emission is usually limited to regions close to the galactic disk due to spectral ageing \citep{Miskolczi_2019}. Hence, the direct mapping of the magnetic field structure and strength in galaxies at higher redshifts will be a challenging and important task for the next generation of radio telescopes, such as the Square Kilometre Array (SKA).

In this paper we use Faraday rotation to examine the magnetic fields around galaxies. Faraday rotation is a process that rotates the polarization angle of linearly polarized light when passing through an ionized and magnetized medium. The intervening medium causes a difference in the phase velocity between the left-handed and right-handed circular polarization components of the linearly polarized synchrotron radiation. Using the polarization angle $\psi$, the observed wavelength $\lambda$ and the Faraday depth $\phi$, the Faraday rotation can be described by a rotation of the intrinsic polarization angle:
\begin{align}
    \psi (\lambda^2) = \psi_0 + \phi \lambda^2.
    \label{rm_equation}
\end{align}
The Faraday rotation measure (RM) modifies the polarization angle via
\begin{align}
\psi (\lambda) = \psi_0 + \rm{RM} \lambda^2.
\end{align}
Faraday rotation of distant background radio sources has been used to probe extragalactic magnetic fields \citep{2022MNRAS.515..256P}. The RM of the polarized light from background radio sources passing through the magnetized CGM can be used as a tracer of the strength and extent of magnetic fields around galaxies out to distances of hundreds of kiloparsecs. The RM is assumed to be positive when the line-of-sight average component of the magnetic field points toward the observer, otherwise it is negative for a field with an average component pointing away from the observer \citep{Vacca_2016}. 
For a polarized radio source at redshift $z$, the RM is defined as
\begin{align}
    \frac{\rm RM}{\rm rad\,m^{-2}} = 0.81 \int\limits_{\rm z}^0 \frac{1}{(1+z)^2} \left (\frac{n_e(z)}{\rm cm^{-3}}\right ) \left ( \frac{B_\parallel(z)}{\mu G} \right ) \left ( \frac {{\rm d}r(z)}{{\rm d}z} \right ) {\rm d}z.
    \label{RM_eq}
\end{align}
The RM has units of \urm, the free electron number density $n_e$ is in cm$^{-3}$, the magnetic field component along the line of sight $B_{\parallel}$ is in Gauss and the comoving path increment per unit redshift $\mathrm{d}l / \mathrm{d}z$ is in parsec. This equation assumes a uniform RM screen across the source and a spatial separation of the linearly polarized source and the Faraday rotating plasma \citep{Akahori_2016, Bernet_2012}. 

Previous work has studied the Faraday rotation properties of background quasars with strong intervening Mg\,{\sc ii} lines in their spectra. Mg\,{\sc ii} absorption is usually associated with the halos of normal galaxies. The absolute values of RM are found to be correlated with the presence of intervening Mg\,{\sc ii} absorption, which is thought to arise in outflowing material from star forming galaxies \citep{Kacprzak_2008, Kim_2016}. Work of \citet{Bernet_2008, Bernet_2010} showed that Mg\,{\sc ii} absorbers exhibit the highest |RM| at frequencies of 5~GHz and above. At lower frequencies, the effect of Faraday depolarization becomes stronger because the Faraday rotation is proportional to the square of the wavelength (\cref{rm_equation}). \citet{Bernet_2013} examined the |RM| distribution with respect to the impact parameters of galaxies finding that all sightlines with high |RM| pass within 50\,kpc of a galaxy and that the |RM| distribution for low impact parameters, $D < 50$ kpc, is significantly different than for larger impact parameters. \citet{Farnes_2014} examined 1.4 GHz data of 599 optically identified non-intrinsic MgII absorption systems with polarized background radio sources finding that the excess of |RM| is still present in that frequency range but only for sources where the impact parameters between the quasar and the polarized emission are small. 

Recently, \citet{Heesen_2023} studied the residual rotation measures (RRMs) observed with the LOw Frequency ARray (LOFAR) around 183 nearby galaxies of the Palomar survey that were selected by apparent $b$-band magnitude \citep{Caretti_2023}. Since Faraday rotation is proportional to the wavelength squared, LOFAR high-band frequencies (144 MHz in this study) afford high-precision RM measurements \citep{OSullivan} at the cost of smaller source densities owing to depolarisation. This work showed, for the first time, an RM along the minor axis of inclined galaxies for impact parameters of less than 100\,kpc. These results suggest a slow decrease of the magnetic field strength with distance from the galactic disc, as expected if the CGM is magnetized by galactic winds and outflows. We note that this work focuses on nearby galaxies that are believed to have a smaller fraction of star-formation driven outflows than at higher redshifts. 

Here, we measure the RM around foreground star-forming galaxies using early-release data from the MeerKAT MIGH-
TEE polarisation survey (MIGHTEE-POL, Taylor et al. 2023, in prep) with the aim to measure the rotation measure profile out to distances of 300--400\,kpc from the star-forming galaxies. This provides information about the magnetic fields in the CGM and galactic winds. We use catalogs of star-forming and blue cloud galaxies to measure the rotation measure of MIGHTEE-POL sources as a function of the impact parameter from the foreground galaxy. We use catalogs of star-forming galaxies since RMs are expected to be higher around star-forming galaxies than around quiescent galaxies. This is due to the fact that the interstellar medium in star-forming galaxies is more turbulent and magnetized, which leads to an increase of the RM. Star formation processes, such as supernova explosions and outflows from young stars, can amplify the magnetic field and drive magnetized outflows on galactic scales \citep{Wiener_2017, Basu_2018}.

This paper is organized as follows: In \Cref{Data_sec} we describe the data and our sample selection and describe the methods in \Cref{Method_sec}. We present the results in \Cref{Results} and discuss them in \Cref{Discussion}. We close with a conclusion in \Cref{Conclusion}.

\section{Data}\label{Data_sec}

The MIGHTEE survey (MeerKAT International GHz Tiered Extragalactic Exploration, \citep{Jarvis_2016}) is a survey that is being conducted using the MeerKAT radio telescope in South Africa. It is one of MeerKAT's flagship Large Survey Projects, using simultaneous continuum, spectropolarimetry \citep{Sekhar_2022} and spectral line \citep{Maddox_2021} measurements to investigate the formation and evolution of galaxies over cosmic time \citep{Heywood_2022}.
The MIGHTEE survey is imaging four extragalactic fields, Cosmic Evolution Survey \citep[COSMOS;][]{Scoville_2007},  {\it XMM} Large-Scale Structure survey \citep[{\it XMM}-LSS;][]{Pierre_2004}, CDFS, and ELAIS-S1. All fields are observed at $L$-band from 880–-1680\,MHz, with a central frequency of 1284\,MHz in multiple pointings that are mosaicked to a final image with a thermal noise sensitivity of approximately 2~$\mu\rm Jy\,beam^{-1}$.
In this work we make use of early-release data products from the MeerKAT MIGHTEE polarisation survey (MIGHTEE-POL; Taylor et al. 2023, in prep). The MIGHTEE-POL survey aims to study the polarized emission from extragalactic radio sources and the properties of magnetic fields in the Cosmic Web including the large-scale structure. The survey will cover an area of about 20 square degrees of the sky, with a resolution of $\approx$ 5 arcseconds. The survey is expected to be completed in 2023 and the data will be made publicly available to the scientific community. In this work, we use continuum data early-release products from the MIGHTEE-POL survey of the COSMOS and \lss\ fields. The visibility data were calibrated and imaged in full polarization mode on the \texttt{ilifu} cloud facility using the CASA-based IDIA pipeline\footnote{https://idia-pipelines.github.io/docs/processMeerKAT}. The analysis of the polarization signals was restricted to frequencies below 1380 MHz in order to minimize instrumental polarization effects. Below this frequency, the instrumental leakage for the MeerKAT remains below $0.2\%$ within 0.5$^{\circ}$ of the field center.

Faraday depth spectra were computed at the location of sources detected in the total intensity catalog by applying the technique of RM synthesis on the fractional Stokes $Q, U$ data between 886 and 1380\,MHz with a channel resolution of 2.51\,MHz. The rotation measure transfer function in these data has a typical width of 56\,\urm\, and the data are sensitive to Faraday depth of up to $\pm5000$\,\urm\ . However, polarized emission was only searched over a Faraday depth range of $\pm$ 2000 \urm\ .
The RM values used in this study were obtained from the location of the strongest polarized component in the Faraday depth spectrum. 
The RM were also estimated by modeling the frequency spectra of Stokes $Q$ and $U$ parameters, and the best-fit RM values were consistent with those obtained from the peak of the Faraday depth spectra within $1\sigma$ (Taylor et al., 2023, in prep). Therefore, we do not expect the values of RM used in this work to be affected by spurious peaks in the Faraday depth spectrum origination due to Faraday complexity \citep{Kim_2016}.

The COSMOS field consists of one pointing, for the \lss\ field three pointings are mosaicked together. The MIGHTEE-POL catalog of the \lss\ field consists of 243 sources and the catalog of the COSMOS field consists of 111 sources. For a more detailed description about the method of data processing see (Taylor et al. 2023, in prep). Compared to lower frequency observations, e.g. with LOFAR, our polarized source density is higher because depolarization is lower. The flip side is that our individual RM values are less accurate as the angle shift depends on $\lambda^2$. In our data the average error on the RM is RM$_{\mathrm{err}}$= 2.4 \urm\ .

\begin{figure}
   \centering
   \includegraphics[width=\linewidth]{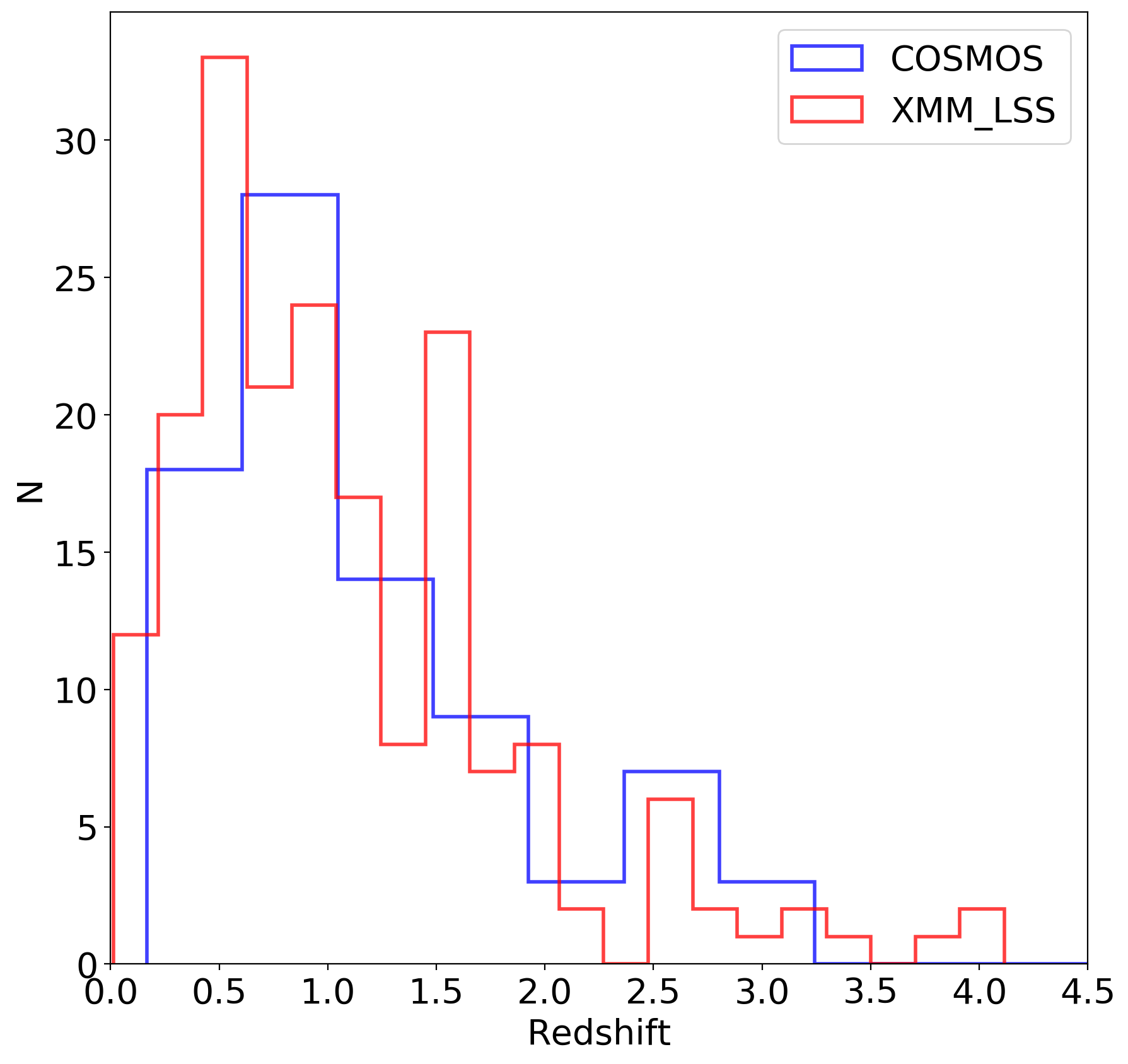}
   \includegraphics[width=\linewidth]{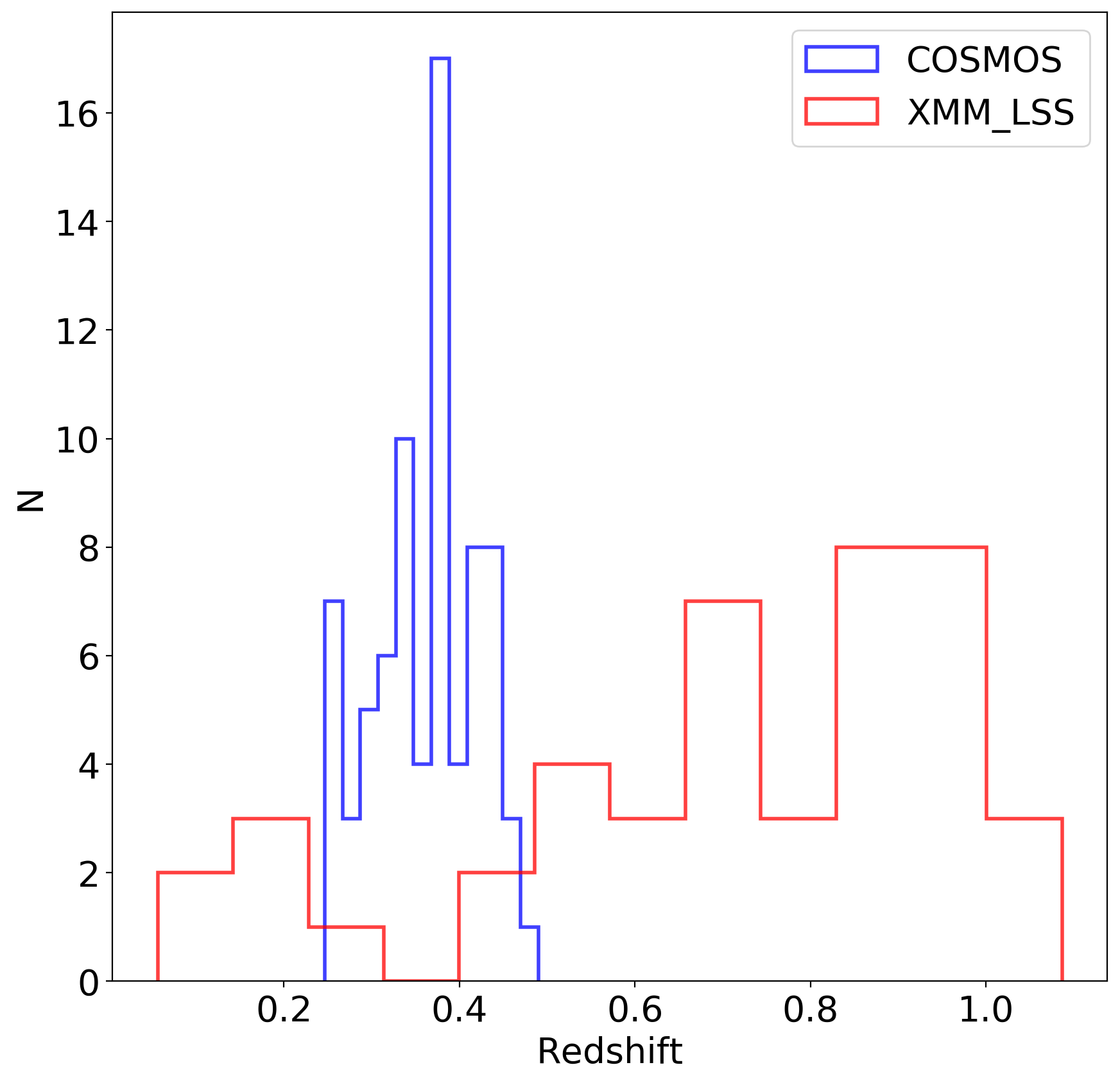}
   \caption{Redshift distributions of host and intervening galaxies. \emph{Upper Panel:\/} The redshift distribution of the host sources in the COSMOS field (blue) and the \lss\ field (red). \\ \emph{Lower Panel:\/} Histogram showing the redshift distribution of the 81 intervening galaxies in the COSMOS field \citep{Sinigaglia_2022} that have matched host sources within a distance of 400kpc in blue and the redshift distribution of the 44 intervening blue cloud galaxies in the \lss\ field \citep{Basu_2015} that have matched host sources within a distance of 400\,kpc in red.}
         \label{redshift_hist_host}
   \end{figure}

\begin{figure}
   \centering
   \includegraphics[width=\linewidth]{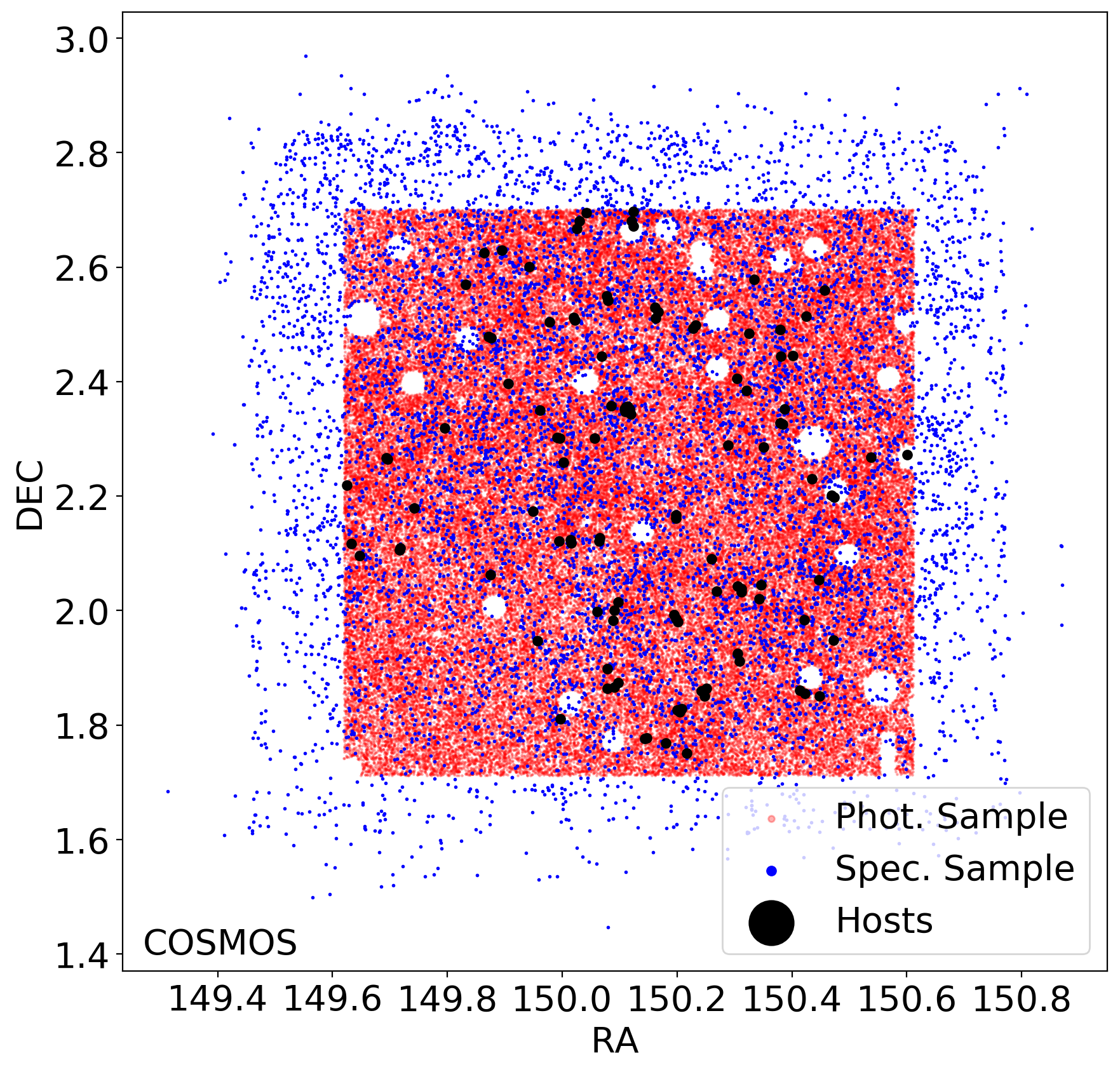}
   \includegraphics[width=9.1cm]{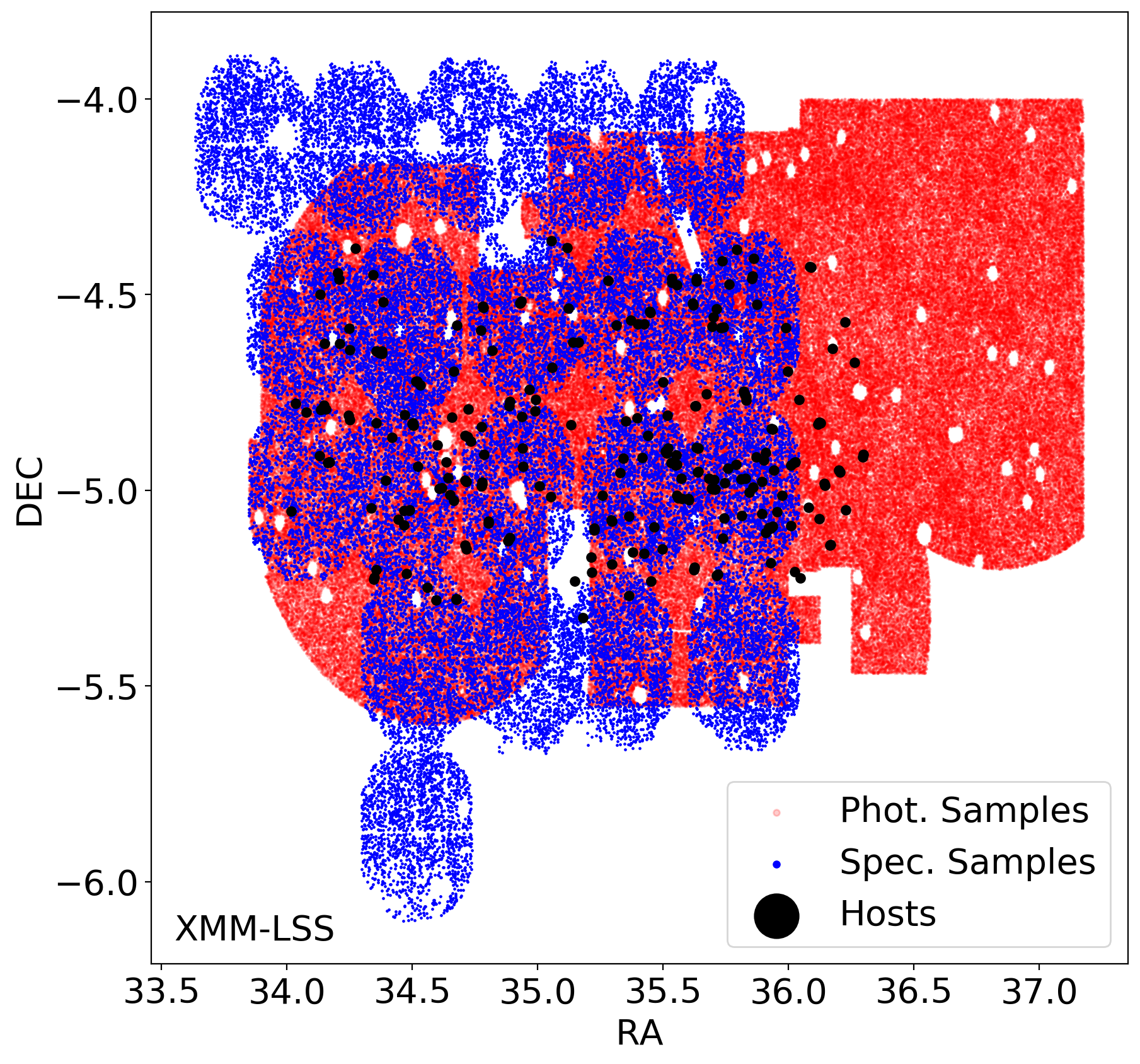}
   \caption{Footprints of all investigated fields. \emph{Upper Panel:\/} We show the distribution of the intervening galaxies for the COSMOS field from the spectroscopic catalog by \citet{Sinigaglia_2022} in blue and the distribution of the corresponding COSMOS MIGHTEE host galaxies in black. The intervening galaxies from the photometric \citep{Hatfield_2022} sample are shown in red.\\
   \emph{Lower Panel:\/} We show the distribution of the intervening galaxies for the \lss\ field from the spectroscopic catalog by \citet{Basu_2015} in blue and the distribution of the corresponding \lss\ MIGHTEE host galaxies in black. The patchy distribution results from the pointings of the Inamori Magellan Areal Camera of the Magellan Telescope at Las Campanas Observatory. The intervening galaxies from the photometric \citep{Hatfield_2022} sample are shown in red.}
         \label{Footprint}
   \end{figure}

The redshift distribution of the host sources is shown in the upper panel of \Cref{redshift_hist_host}. In order to identify intervening galaxies in the \lss\ field  we use the "blue cloud" galaxy catalog from \citet{Basu_2015}. This catalog consists of 36776 blue galaxies with their spectroscopic redshifts from the PRism MUltiobject Survey (PRIMUS) up to a redshift of $z = 1.18$. A color-magnitude diagram based separation was introduced to only select blue star-forming galaxies. Furthermore, the PRIMUS team identified the AGN in their sample by fitting AGN spectral templates to remove AGN from the normal galaxy sample. In this catalog, due to the evolving main-sequence, high-$z$ blue cloud galaxies tend to be luminous infrared galaxies (LIRGs) and ultra-luminous infrared galaxies (ULIRGs), and therefore dominated by mergers.\\
For the COSMOS field we use the galaxy catalog provided by \citet{Sinigaglia_2022}. This sample was derived from the parent sample by \citet{Weaver_2022}. A color–color $NUV - r/r - J$ plane selection was applied to only select star-forming galaxies. This catalog contains 9022 star-forming galaxies and their spectroscopic redshifts between $0.23 < z < 0.48$. Note that this sample is the result of the combination of several different surveys, performed with different survey strategies \citep{Weaver_2022, Davies_2018}. 
The distribution of the locations of intervening and host galaxies across the sky for both fields is shown in \Cref{Footprint}. We can see that the optical survey used in the catalog creation for the \lss\ field has a smaller footprint than the MIGHTEE survey. This reduces the size of the \lss\ field.

In addition to the two spectroscopic galaxy catalogs we also make use of two photometric galaxy catalogs (one for each field) provided by \citet{Hatfield_2022}. These catalogs were computed with optical and near-infrared data from VISTA and HyperSuprimeCam (VISTA Deep Extragalactic Observations, VIDEO: \citet{Jarvis_2013}; HSC: \citet{Aihara_2018}). We remove sources without redshift and magnitudes and implement a cut on the apparent Ks magnitude of $m_K < 23.7$. The magnitude cut is introduced to ensure uniformity and comparability between the two catalogs, given that the initial depth of the COSMOS catalog exceeds that of the \lss\  catalog by 1 magnitude. After doing so, we are left with 125676 sources for the COSMOS field and 384354 sources for the \lss\ field. The redshift range for the COSMOS field is $0.001 < z < 5.65$ with a median $g$-band magnitude of $M_g = -19.1\,\rm mag$. For the \lss\ we get $0.0001 < z < 6.45$ a median $g$-band magnitude of $M_g = -20.3\,\rm mag$. We use these catalogs to find the total number of intervenors to each host galaxy and to test for a correlation between the mean impact parameter of all intervenors and |RM| as these catalogs are more complete than the pure spectroscopic samples. Also, these catalogs provide the mass and star-formation rate of each galaxy which we use for our analysis. We note that the photometric galaxy catalogs include subsets of the spectrocopic catalogs. However, these catalogs do not overlap completely, which is visible in \Cref{Footprint}. 
Because the galaxies in the spectroscopic catalogs are chosen with certain selection criteria, it is possible that they are in general incomplete. In order to tackle this, we considered a more comprehensive sample by additionally using photometric catalogs.

\section{Method}\label{Method_sec}

\subsection{The spectroscopic sample}

In order to measure the RM induced by the CGM of the intervening galaxies, we need to associate the background sources with the intervening sources. For each background source, we search for an intervening galaxy within a projected distance of less than 400 kpc at the redshift of the intervening source. If more than one intervening galaxy matches a host source, we chose the one with the minimum central distance. We only select sources where $z_{\mathrm{Host}} > z_{\mathrm{Intervenor}}$ to ensure that the intervening galaxy lies in front of the host galaxies. For the \lss\ galaxy sample, we introduce a magnitude cut and exclude galaxies with $M_u > -22$\,mag.

The $u$-band magnitude is an indicator for star-forming galaxies as the UV light is primarily emitted by hot, young, massive stars \citep{Calzetti}. $M_u$ and $M_g$ are comparable and only differ by a value of 0.4\,mag on average. Our magnitude cut does not impose any constraint on the initial catalog as the spectroscopic survey goes down to apparent magnitudes of $m_g = 27$ and an absolute magnitude of $M_g = -22$ converts to an apparent magnitude of $m_g = 23.1$ at redshift $z = 1$. This leads to the result that no galaxies in our magnitude range are cut from the catalog. In addition, by introducing the cut we select more massive galaxies which leads to more massive galactic halos which in turn may lead to a denser environments with a larger RM.

By matching the catalogs and after removing duplicates we are left with 81 (44) matches for the COSMOS (\lss) spectroscopic samples, see \Cref{specs}. As we have more volume to detect brighter galaxies at the higher redshift we add a bias by introducing a magnitude cut to our intervening galaxy sample due to the shape of the luminosity function. The redshift distributions of both samples of the intervening galaxies are shown in the lower panel of \Cref{redshift_hist_host}.

We do not correct the RM values for Galactic contamination by the Milky Way because we only look at very small patches on the sky, $1.6\,\rm deg^2$ for the COSMOS field and $3.5\,\rm deg^2$ for the \lss\ field. Hence, we expect a negligible gradient over these small sky patches and assume the contribution of the Milky Way for each field is constant over all sources and therefore not affecting our analysis. Nonetheless, we note that (Taylor et al. 2023, in prep) found the median Galactic contribution (Galactic Rotation Measure; GRM) of the \lss\ field to be $8.9\pm 3.7$\,\urm\ and of the COSMOS field to be $0.9\pm 4.1$\,\urm. We note that we find fewer matches for the \lss\ sample, although the \lss\ field is larger than the COSMOS field. This is due to the introduced magnitude cut which reduces the numbers of matches from 168 to a number of 44 which is further discussed in \Cref{Discussion}.

\begin{table*}
\begin{center}
\caption[]{This table lists the properties of the background radio sources and intervenors in the spectroscopic galaxy catalogs in the COSMOS and \lss\ field. $N_{\mathrm{host}}$ refers to the total numbers of host galaxies from the MIGHTEE sample, $N_{\mathrm{spec}}$ are the total numbers of galaxies in the spectroscopic catalogs, $N_{\mathrm{matched}}$ are the numbers of matches for the hosts, $z_{\mathrm{spec}}$ is the redshift range of the galaxies in the spectroscopic catalogs and reference is the reference paper.}
\label{Sample}
\begin{tabular}{llllll}
\hline
Field & $N_{\mathrm{host}}$ & $N_{\mathrm{spec}}$ & $N_{\mathrm{matched}}$ & $z_{\mathrm{spec}}$ & Reference \\
\hline
COSMOS & 111 & 9,022 & 81 & 0.25--0.49 & \citet{Sinigaglia_2022}\\
\lss\ & 243 & 36,776 & 44 & 0.06--1.09 & \citet{Basu_2015}\\
\hline
\end{tabular}
\label{specs}
\end{center}
\end{table*}

For both of our samples, we calculate the impact parameter which is the separation between the background source and the center of the foreground galaxy in kpc. For our analysis we use the absolute value of the rotation measure, |RM|. Special care needs to be taken to compute the errors because the |RM| are an absolute quantity.
The error on |RM| was calculated by assuming that the error on RM follows a Gaussian distribution. For each value of RM, we generated a random sample of |RM| drawn from an underlying Gaussian distribution, with a mean value of RM and standard deviation given by the measured error on RM. We then consider the 68\,percentile ($\equiv 1\,\sigma$) interval of the distribution of |RM|, centered at the median value of the distribution, as the error on each |RM|.

\begin{figure}
   \centering
   \includegraphics[width=\linewidth]{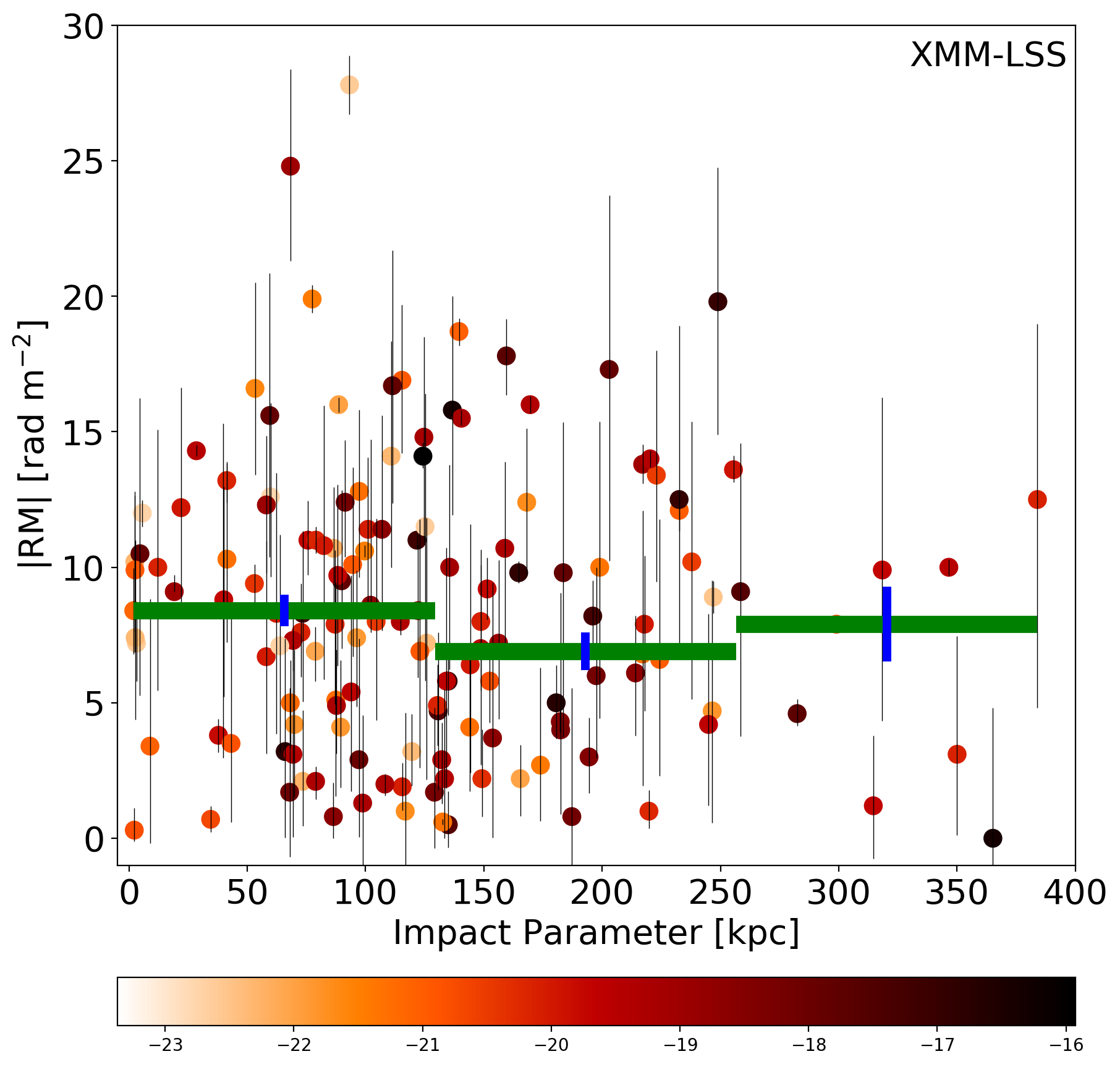}
   \caption{Observed |RM| in \urm\ versus impact parameter in kpc for the \lss\ without the magnitude cut. The green bar indicates a bin of individual data points with the blue line being the error bar of the bin. The individual data points are color-coded with their corresponding $u$-band magnitude.}
         \label{XMM_no_cut}
   \end{figure}

In order to investigate the excess of |RM| at smaller impact parameters, we bin the data points. For the typical mass of our galaxies, we expect a virial radius of $\approx$150--200\,kpc. Thus, we compute three bins with an approximately equal number of objects, each with a width of 133\,kpc in impact parameter. In each bin we calculate the median values of |RM| and follow the procedure of \citet{Aramburo-Garcia_2023} to estimate the error in each bin using:
\begin{align}
|{\rm RM}|_{\mathrm{err, bin}} = \sqrt{\frac{\left\langle \left( x_i - \langle x \rangle \right)^2 \right\rangle}{n}} ,
\label{bin_error}
\end{align}
where $x_i$ are |RM| values in each bin, $\langle x \rangle$ is their median value and $n$ is the number of objects in each bin. 
We then plot the observed |RM| in \urm\ versus the impact parameter in kpc for both samples. First, we show the observed |RM| in \urm\ versus the impact parameter for the \lss\ without the magnitude cut of $M_u > -22$\,mag of that we introduced in order to mainly select massive star-forming galaxies. As we can see in \Cref{XMM_no_cut}, this increases the number of matches of intervenors and hosts for the sample to 168 but we do not detect any excess if we also include faint galaxies with $M_u > -22$\,mag. 

\subsection{The photometric sample}

Using the spectroscopic sample, we only investigate the RM contribution from one massive, bright intervenor with the smallest impact parameter to the host. However, as the  RM is an integral along the line of sight to the background radio source, we also investigate how the number of intervenors $N_{\rm int}$ and the mean impact parameter between a host and all its intervenors varies with the total RM of each host. The spectroscopic galaxy catalogs that we use for our analysis are incomplete in terms of magnitude and redshift and are only a subset of galaxies that lie between host and observer. In order to examine the total numbers of intervenors to each host we make use of the photometric catalogs from \citet{Hatfield_2022} that contain more than 100.000 galaxies for each field to identify all intervenors to each host within a certain impact parameter.

We identify all intervening galaxies within an impact parameter of 133 kpc for each host galaxy in the two samples. We do not apply any cuts on magnitude or star-formation rate. For every host galaxy we compute the mean impact parameter to all its intervenors to investigate if there is a connection to the total |RM|. In addition, we recognize that galaxies with higher rates of star-formation are expected to exhibit stronger magnetic fields. Thus, we assume that these galaxies contribute more significantly to the total |RM| signal. We introduce a weighting scheme with respect to the impact parameter to account for the correlation between impact parameter and |RM|. The introduced weighting scheme allows us to explicitly account for the relative importance of each galaxy in the final analysis. The weighting is introduced as follows
\begin{align}
\mathrm{IP}_\mathrm{{weighted,mean}} = \frac{\sum_i  \mathrm{SFR}_i \, \mathrm{IP}_i}{\sum_i \mathrm{SFR}_i},
\label{weighting}
\end{align}
where SFR$_i$ is the star-formation rate and IP$_i$ is the impact parameter in kpc of the individual intervening galaxy. Given the fact that we obtain individual impact parameters for every intervenor to the corresponding host and we only obtain a single observable value of RM which remains the same for every host, the weighting strategy must be implemented for the impact parameters. This is necessary to ensure that each individual impact parameter is appropriately considered in the overall analysis, taking into account its unique contribution to the final result.

\section{Results}\label{Results}

\subsection{The spectroscopic sample}

\begin{figure}[h!]
   \centering
   \includegraphics[width=\linewidth]{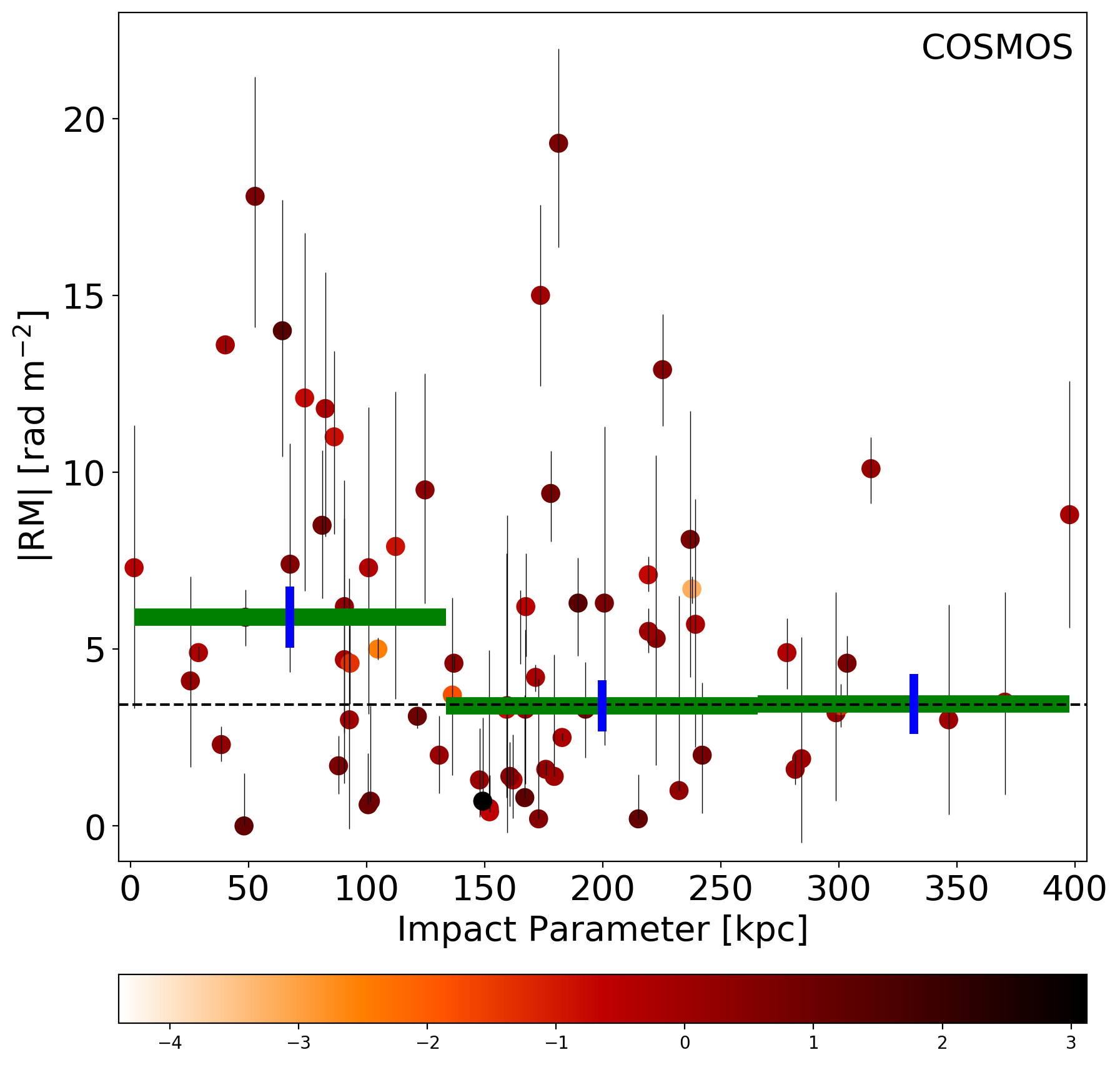}
   \includegraphics[width=\linewidth]{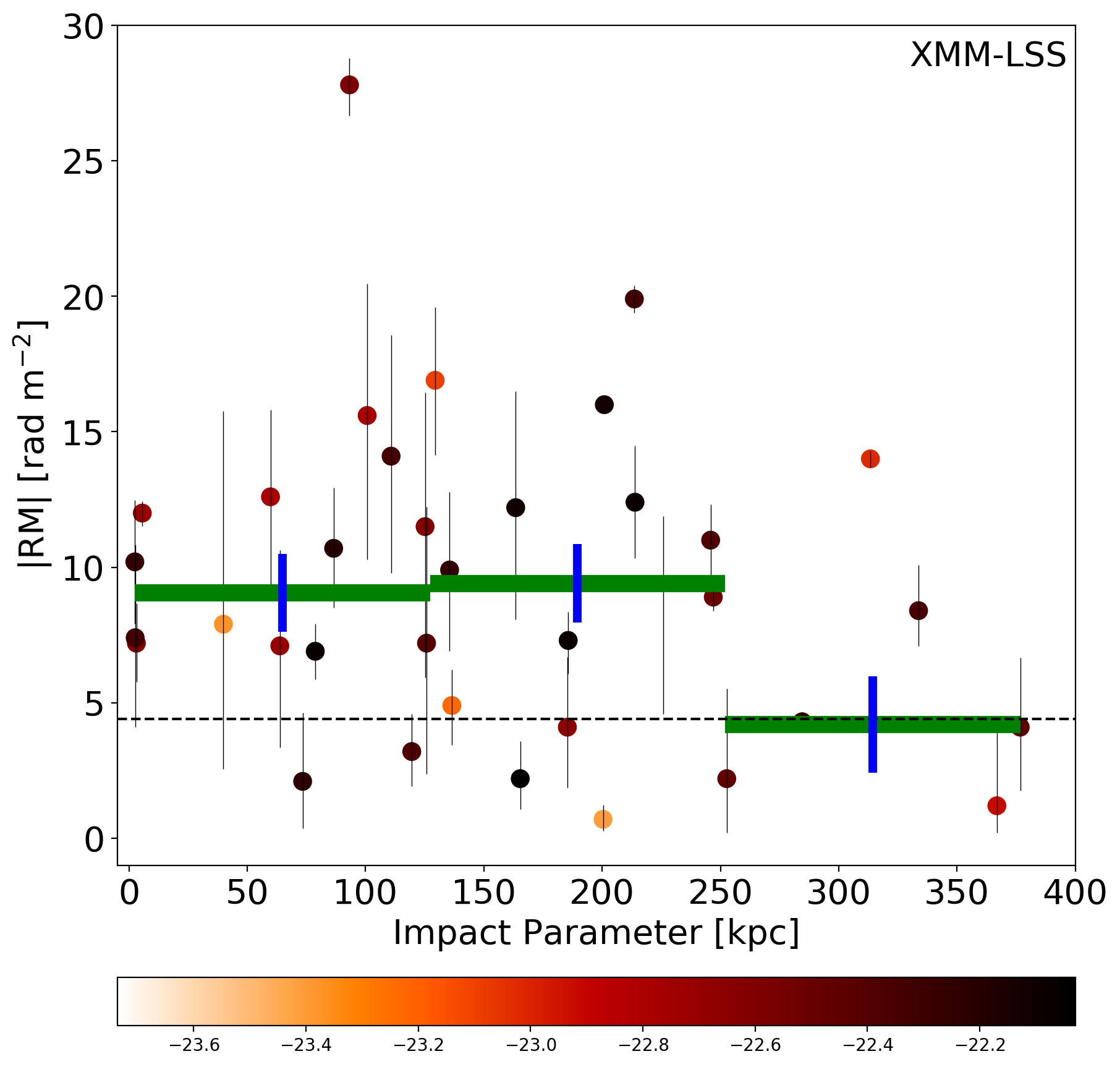}
   \caption{Excess |RM| as function of impact parameter for the spectroscopic samples. \emph{Upper Panel:\/} Observed |RM| in \urm\ versus impact parameter in kpc for all 81 sources in the COSMOS field. The green bar indicates a bin of individual data points with the blue line being the error bar. The data points are color-coded with respect to the star-formation rate of each galaxy. \emph{Lower Panel:\/} Observed |RM| in \urm\ versus impact parameter in kpc for all 44 sources in the \lss\ field. Here, the data points are color-coded with respect to the $u$-band magnitude of each galaxy. For both fields all error bars show the 68 percentile interval around |RM|.}
         \label{RM_impact}
   \end{figure}

\Cref{RM_impact} shows the resulting plots for the spectroscopic COSMOS and \lss\ samples after performing the magnitude cut. For the COSMOS sample we are provided with the star formation rate (SFR), we include this information in the plot and color-code the individual data points with the SFR. For the \lss\ we have the individual magnitudes and we use this information in the corresponding plot for color coding the individual data points. 

We calculate the absolute RM excess \emph{un}corrected for the intervening galaxies redshift. The total measured |RM| is a combination of GRM, extragalactic and noise components. The extragalactic |RM| consists of contributions from the invervening galaxies as well as from the foreground IGM between the source and the Milky Way as well as from contributions from the cosmic web. 

For the COSMOS field that consists of a sample size of 81 we find a median |RM| of $5.9\pm 0.9$\,\urm\ for sources with impact parameters less than 133\,kpc and $3.4\pm 0.7$\,\urm\ for sources with impact parameters greater than 133\,kpc. This results in an excess of $2.5\pm 1.1$\,\urm, which is significant at 2.3$\sigma$. For the \lss\ field  that consists of a sample size of 44, the median |RM| was found to be $9.2\pm 1.4$\,\urm\ for sources with impact parameters less than 266\,kpc and $4.2\pm 1.8$\,\urm\ for sources with impact parameters greater than 266\,kpc. This corresponds to an excess of $5.0\pm 2.3$\,\urm, which is significant at 2.2$\sigma$. The median |RM| for high impact parameters of  3.4 and 4.2\,\urm\ with the median of the absolute value being 0.67 of the standard deviation for a Gaussian distribution. These numbers are consistent with results for the standard deviation of extragalactic sources, corrected for the Milky Way of $\sigma_{\mathrm{RM}}= 6$\,\urm\ found by \citet{Schnitzeler_2010}. 

\begin{figure}
   \centering
   \includegraphics[width=\linewidth]{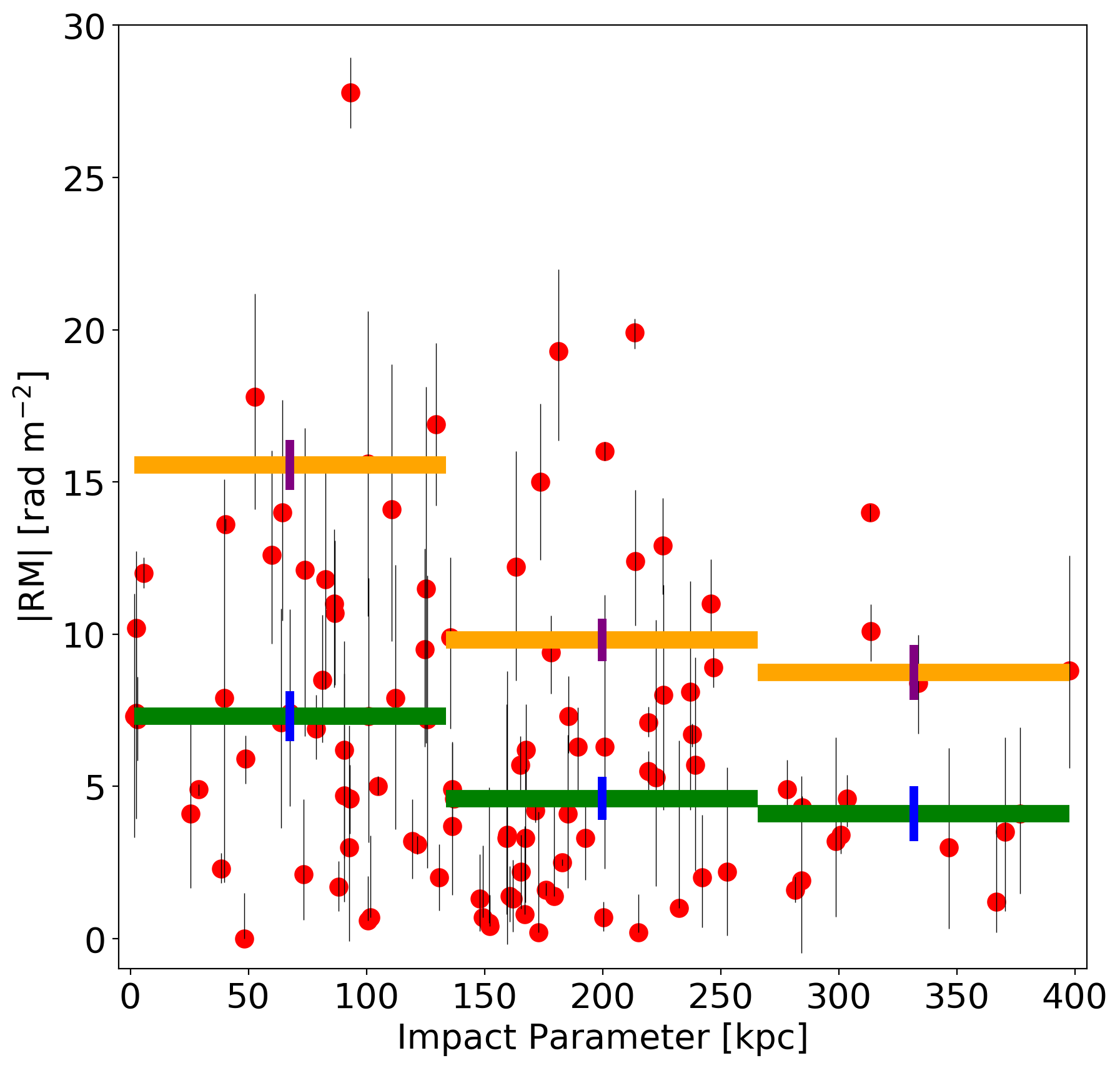}
   \caption{Excess |RM| as function of impact parameter for the spectroscopic sample, combined for the COSMOS and \lss\ field. Observed |RM| in \urm\ versus impact parameter in kpc for the combination of the two samples. The green bar indicates a bin of individual data points with the blue line being the error bar of the bin. The orange bins are the redshift corrected |RM| including the error bar in purple. The error bars show the 68 percentile interval around |RM|.}
         \label{RM_impact_COMBINED}
   \end{figure}

When the samples from both fields are combined, as shown in \Cref{RM_impact_COMBINED}, the median |RM| is found to be $7.3 \pm 0.8$\,\urm\ for sources with impact parameters less than 133\,kpc and $4.3\pm$0.9\,\urm\ for sources with impact parameters greater than 133\,kpc. This results in an excess of $3.0\pm 1.2$\,\urm, which is significant at 2.5$\sigma$. 

As we do not have information on the individual contributions, a redshift correction would boost all contributions although they do not all occur at the same redshift. Hence, we use the redshift uncorrected measurements to investigate the |RM| in correlation with impact parameters and only perform a redshift correction on our binned results using the mean redshift of the intervenors. 

We correct the |RM| excess that we found for redshift effects to get the rest frame RM:
\begin{equation}
    \mathrm{RM}_\mathrm{corr} = \mathrm{RM} (1 + z_i)^2 ,
\end{equation}
with $z_i$ being the redshift of the intervenor. The median redshift for the COSMOS sample is $\bar{z}_\mathrm{COSMOS} = 0.37$ and for the \lss\ sample it is $\bar{z}_\mathrm{XMM-LSS} = 0.77$. If we correct the |RM| excess for redshift we now get $4.7\pm 2.1$\,\urm\ for the COSMOS field and $15.7 \pm 7.2$\,\urm\ for the \lss\ field. For the combination of both samples we find a median redshift of $\bar{z}_\mathrm{Comb} = 0.42$ yielding to a redshift-corrected |RM| excess of $5.6\pm 2.3$\,\urm. The redshift correction of the bins is also shown in \Cref{RM_impact_COMBINED}. A two-sample Kolmogorov--Smirnov test yields a $p$-value of 0.01, which implies a clear significance for the excess. 

\begin{figure}
   \centering
   \includegraphics[width=\linewidth]{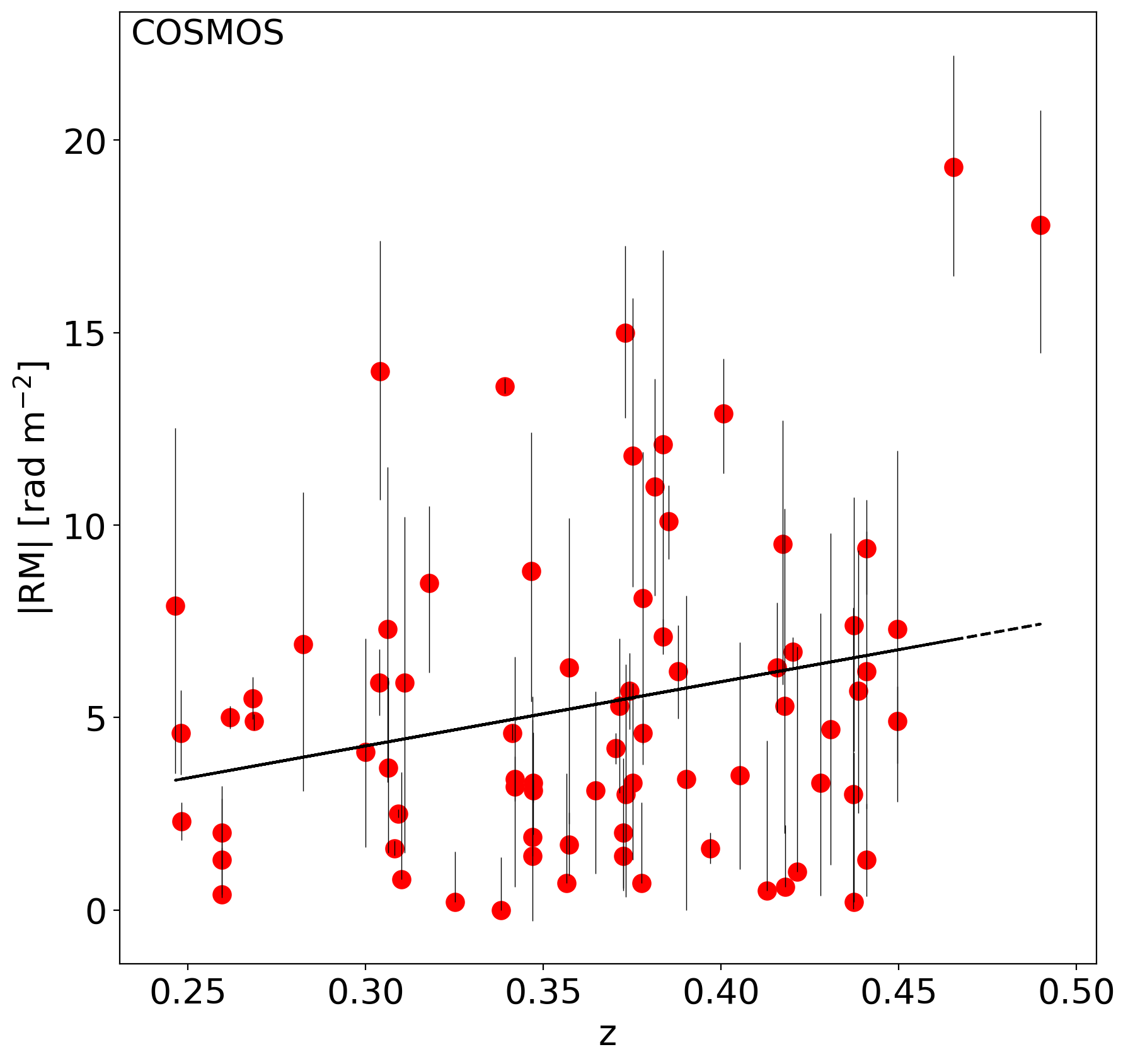}
   \includegraphics[width=\linewidth]{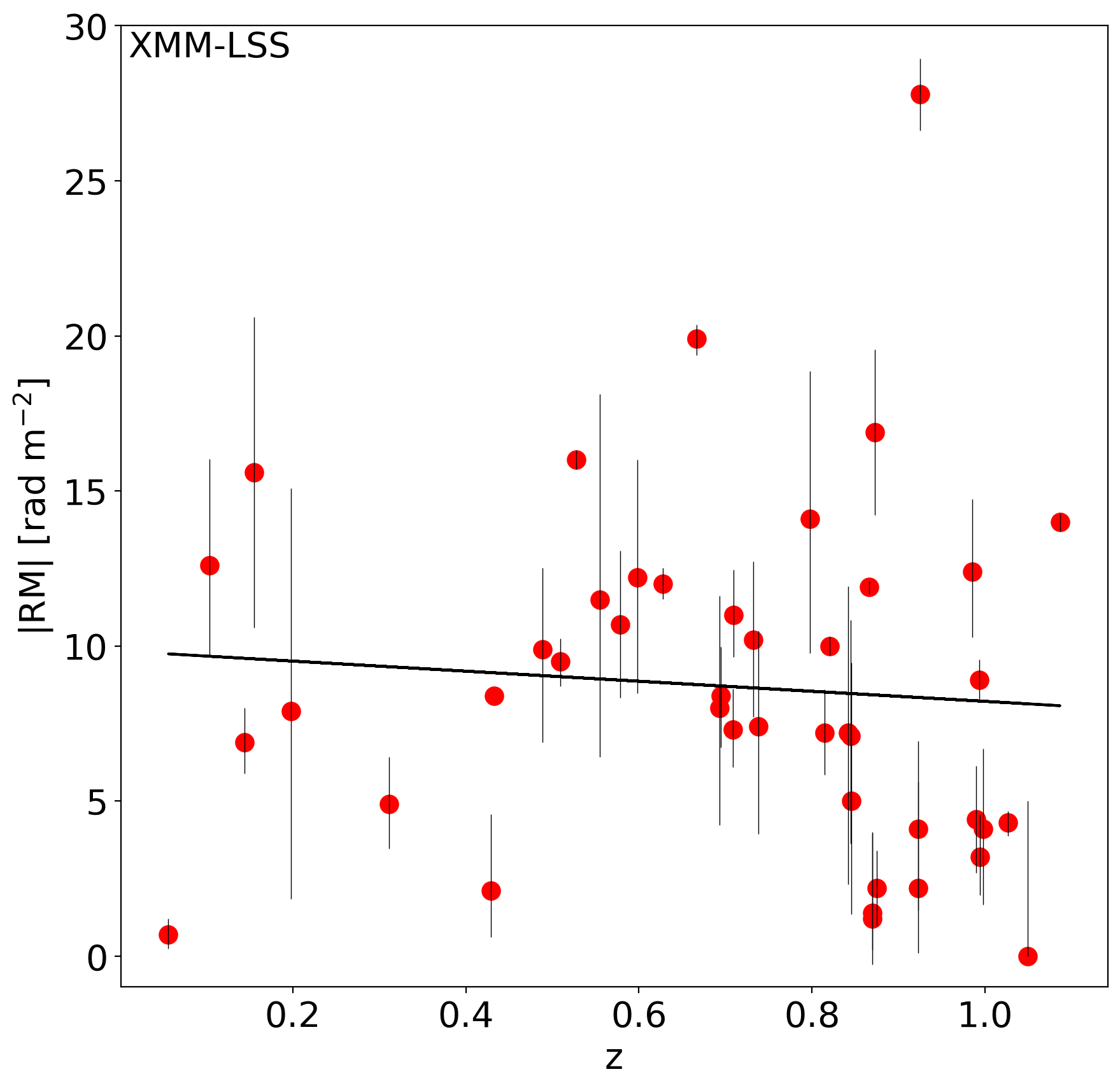}
   \caption{|RM| as function of redshift for the spectroscopic sample. \emph{Upper Panel:\/} Observed |RM| in \urm\ versus redshift for all 81 sources in the COSMOS field. The black line represents the best fit including all data points. \emph{Lower Panel:\/} Observed |RM| in \urm\ versus redshift for all 44 sources in the \lss\ field. Again, the black line represents the best fit. For both fields all error bars show the 68 percentile interval around |RM|.}
         \label{RM_redshift}
   \end{figure}

We also investigate a possible connection between the |RM| of the background source and the redshift of the intervening galaxy. To this end, we compute a linear fit between |RM| and redshift. \Cref{RM_redshift} shows the |RM| vs. the redshift plot of each sample and the fit. In order to evaluate this correlation statistically, we perform a fit on the data and calculate the Pearson correlation coefficient to evaluate the linear correlation between the data and the fit. 
For the COSMOS data we get a positive trend with higher |RM| at higher redshifts and a correlation coefficient of $0.59$ with a $p$-value of $0.12$, suggesting that this trend is not significant. In contrast, the results for the \lss\ field indicate a negative trend with a declining |RM| towards higher redshifts. However, we note that the Pearson correlation coefficient yields $0.05$ for this sample with a $p$-value of $0.9$. So we do not detect a correlation between |RM| and redshift for the \lss\ but we see an indication for a positive correlation for the COSMOS sample.

\subsection{The photometric sample}

\begin{figure}
   \centering
   \includegraphics[width=\linewidth]{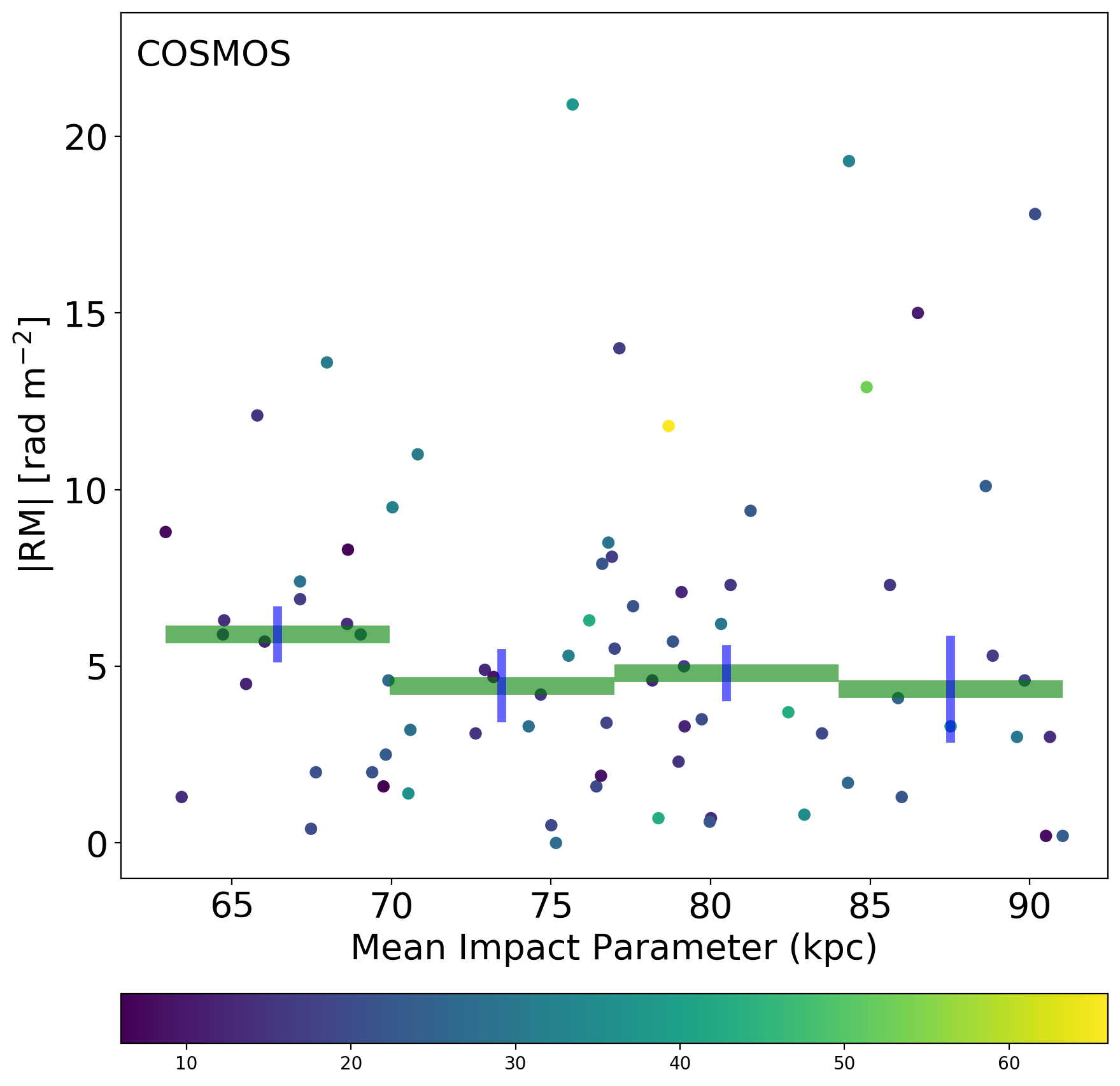}
   \includegraphics[width=\linewidth]{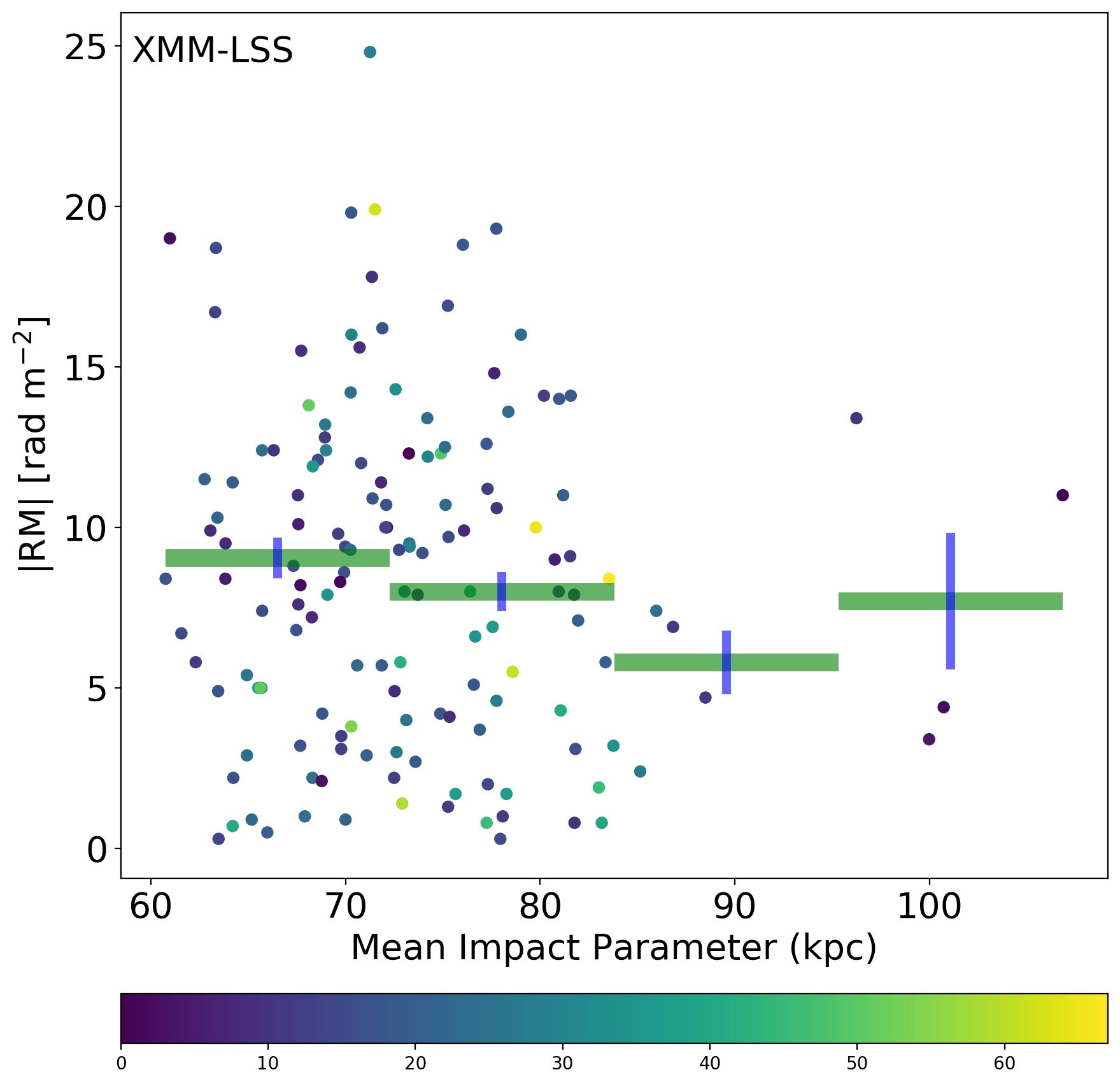}
   \caption{|RM| as function of mean impact parameter for the photometric sample. \emph{Upper Panel:\/} We plot the observed |RM| in \urm\ versus the mean impact parameter of all intervening galaxies within 133kpc around the host from the photometric catalog provided by \citep{Hatfield_2022} for the COSMOS field. The green bar indicates a bin of individual data points with the blue line being the error bar of the bin. The individual data points are color-coded with respect to the total number of intervenors to each host. \emph{Lower Panel:\/} The corresponding plot for the \lss\ field.}
         \label{RM_unweighted}
   \end{figure}

We plot the mean impact parameter of each host to all its intervenors versus |RM| treating all intervenors equally in \Cref{RM_unweighted}. In addition, we color-code the individual data points with respect to the total number of intervenors $N_{\rm int}$ to each host to investigate a possible correlation between $N_{\rm int}$ and |RM|. The number of intervenors to each host within an impact parameter of 133 kpc reach from $6 < N_{\rm int} < 66$ with the median being $\langle N_{\rm int}\rangle$ = 20 for the COSMOS field, and 6 < $N_{\rm int}$ < 67 with the median being $\langle N_{\rm int} \rangle$ = 18 for the \lss\ field. The total numbers from the photometric catalog are presented in \Cref{Hatfield_Cat}. We investigate the correlation between the mean impact parameter to the total |RM| by binning the data. Again, we calculate the error of each bin as it is described in \Cref{bin_error}. Inspecting the data in terms of $N_{\rm int}$ does not yield any correlation between the total number of intervenors $N_{\rm int}$ to the total |RM|. 
Binning the data does not lead to definitive results or trends for either sample. However, caution is necessary when interpreting the data, particularly in cases where the bins with larger impact parameters include only few data points. The absolute magnitudes of the intervenors are in the range of $-20.1\,{\rm mag} < M_g < -12.1\,{\rm mag}$ with a median of $\langle M_g\rangle = -15.5$\,mag for the \lss\ field and of $-18.9\,{\rm mag} < M_g < -13.8\,{\rm mag}$ with a median of $\langle M_g\rangle = -14.8$\,mag for the COSMOS field. This means that the galaxies in the photometric sample go down to much lower brightness than in the spectroscopic sample with a magnitude cut in the $u$-band. Even though the photometric catalogs includes most of the galaxies from the spectroscopic samples, including a large number of faint galaxies to contribute to the RM leads to a dilution of the central RM excess seen for the bright galaxies in the spectroscopic sample.

Evidently, the photometric catalog extends to larger redshifts than the spectroscopic catalog. Since we use the redshift of each intervenor to calculate the impact parameter, we also propagate the error to the impact parameter calculation. The mean redshift of the intervenors in the two catalogs is $\langle z_{\rm int} \rangle = 0.26$ with a mean error on the photometric redshifts of $\langle z_{\rm err} \rangle \sim 0.08$ \citep{Hatfield_2022} which yields an impact parameter error of $\pm 30$ kpc on average. 

\begin{figure}
   \centering
   \includegraphics[width=\linewidth]{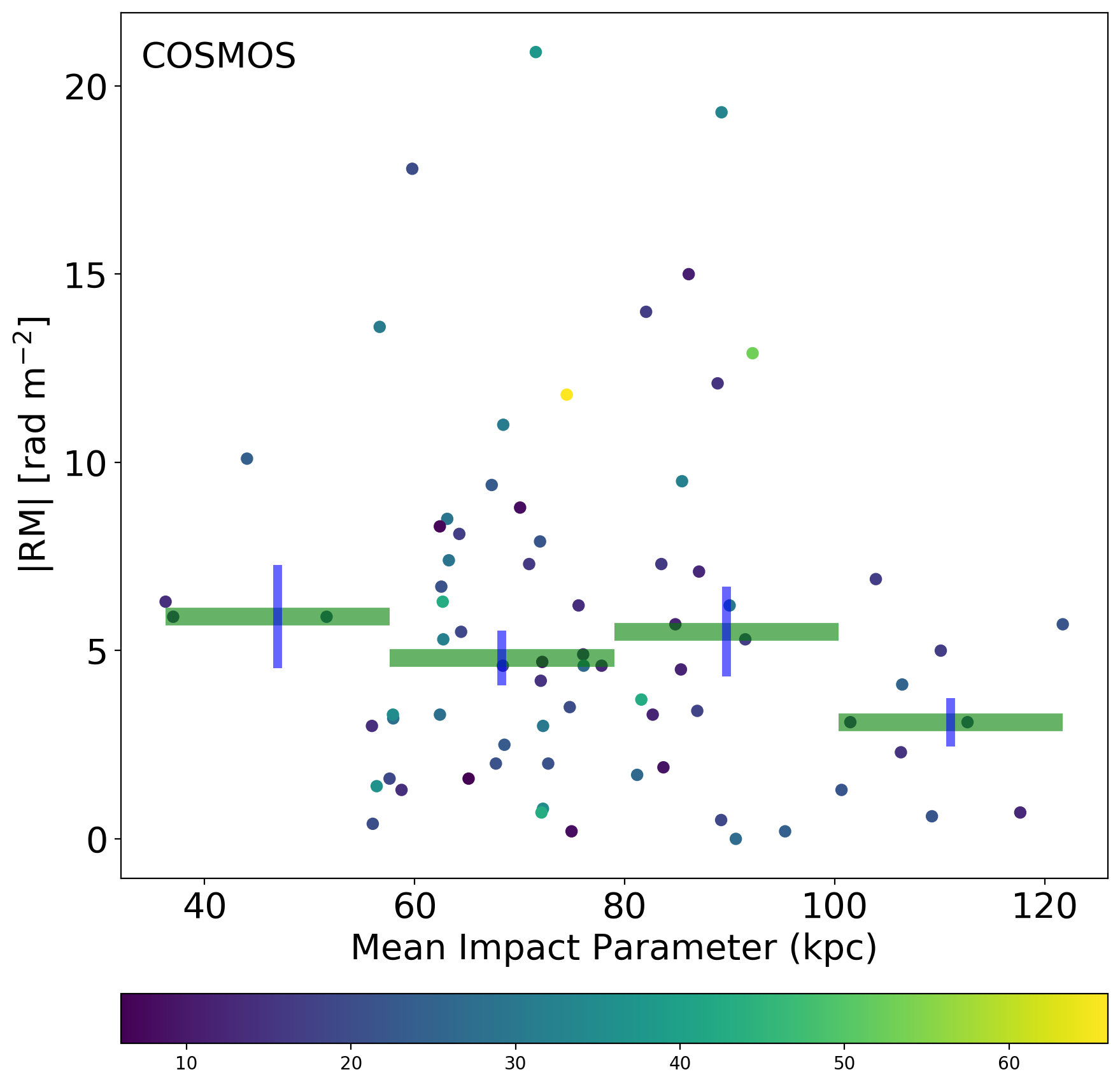}
   \includegraphics[width=\linewidth]{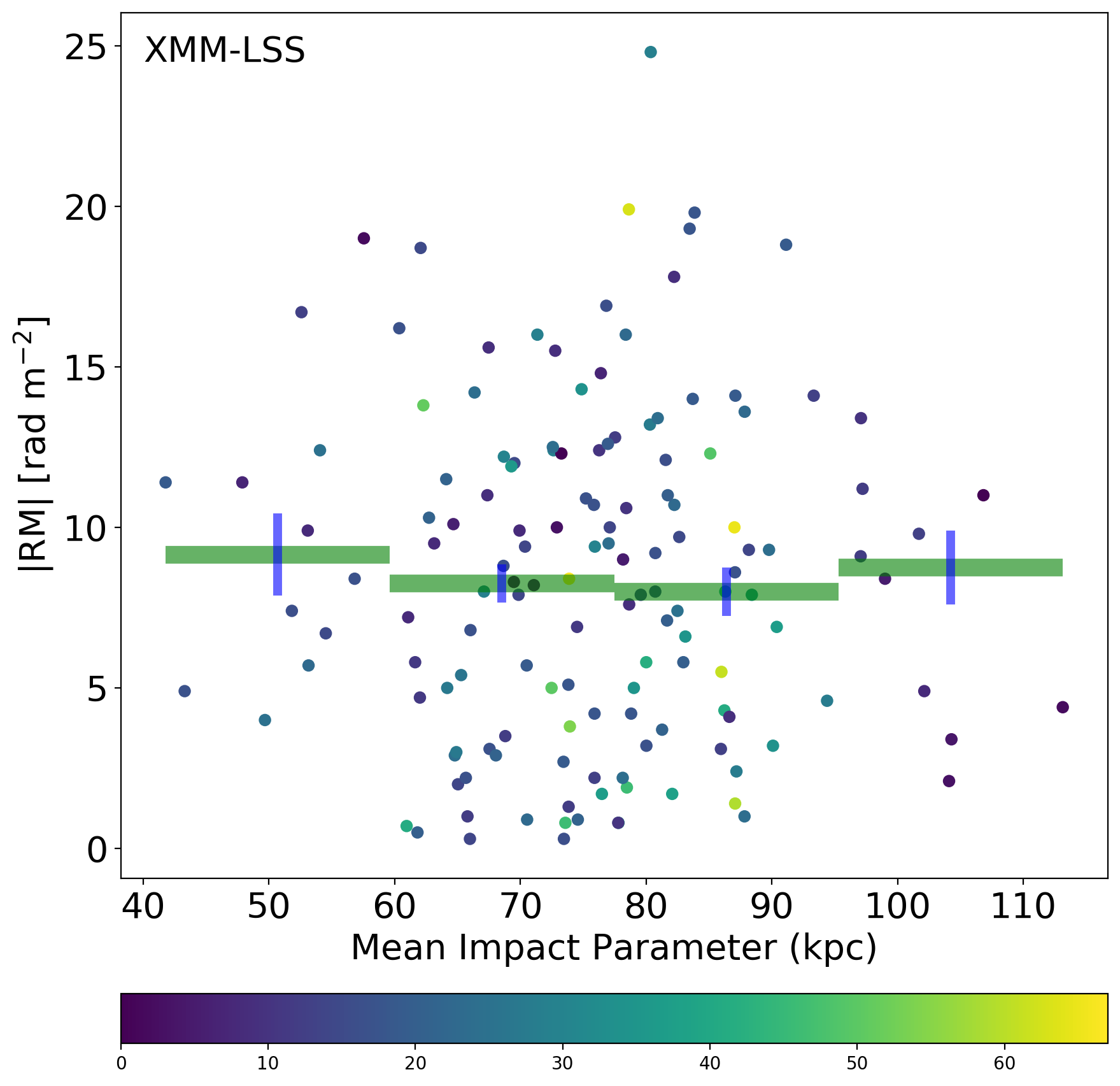}
   \caption{|RM| as function of weighted mean impact parameter for the photometric sample. \emph{Upper Panel:\/} We plot the observed |RM| in \urm\ versus the weighted median impact parameter in kpc from the photometric catalog provided by \citep{Hatfield_2022} within an impact parameter of 133\,kpc around each host for the COSMOS field. The green bar indicates a bin of individual data points with the blue line being the error bar of the bin. The individual data points are color-coded with respect to the total number of intervenors to each host. \emph{Lower Panel:\/} The corresponding plot for the \lss\ field. }
         \label{RM_combined}
   \end{figure}

Next, we re-compute the mean impact parameter of each host to all its intervenors versus the total |RM| applying a weighting by the star-formation rate introduced in \Cref{weighting}. The weighting scheme takes into account the intervenors' star-formation rate and we assume that intervenors with higher star formation rates contribute more to the |RM|. In \Cref{RM_combined} we show the resulting plot. Again, in both samples, we do not find any excess of |RM| at smaller impact parameters.

\begin{table*}
\begin{center}
\caption[]{This table lists the properties of the photometric galaxy catalog from \citet{Hatfield_2022} and the median properties of the intervenors we find to the host sample. $N_{\mathrm{host}}$ are the total numbers of host galaxies from the MIGHTEE sample, $N_{\mathrm{photometric}}$ are the total numbers of galaxies in the photometric catalog for each field, $\langle N_{\rm int}\rangle$ is the median number of intervenors for each host, $\langle z_{\rm phot,int} \rangle$ is the median redshift of the intervenors, $\langle M_{g, \mathrm{int}}\rangle$ is the median $M_{g}$ magnitude of the intervenors, and $\langle m_{\mathrm{int}}\rangle$ and $\langle SFR_{\mathrm{int}} \rangle$ are the median mass and star-formation rate of all intervenors in $\log_{10} (M_{\odot}$) and $\log_{10} (M_{\odot}$/yr) respectively. }
\begin{tabular}{llllllll}
\hline
Field & $N_{\mathrm{host}}$ & $N_{\mathrm{photometric}}$ & $\langle N_{\rm int}\rangle$ & $\langle z_{\rm phot,int} \rangle$ & $\langle M_{g, \mathrm{int}}\rangle$ (mag) & $\langle m_{\mathrm{int}}\rangle$ $\log_{10} (M_{\odot}$)& $\langle SFR_{\mathrm{int}} \rangle$ $\log_{10}(M_{\odot}$/yr)\\
\\
\hline
COSMOS & 111 & 125,676 & 20 & 0.26 &-14.8 & 8.1 & -10.1\\
\lss\ & 243 & 384,354 & 18 & 0.22 &-15.5 & 7.8 & -12.2  \\
\hline
\end{tabular}
\label{Hatfield_Cat}
\end{center}
\end{table*}

\section{Discussion}\label{Discussion}

\subsection{The spectroscopic sample}

We detect of an excess of |RM| for small impact parameters in both MIGHTEE fields. The COSMOS sample shows an excess at a radius below 133\,kpc around the galaxy whereas the \lss\ exhibits an excess up to a radius of 266\,kpc. The higher median redshift of the \lss\ sample would suggest that also the virial radius of these galaxies is larger as we are biased to more massive systems at higher redshift which is in agreement with our results. We note that we detect the excess in the \lss\ sample only if we introduce a magnitude cut for intervenors with $M_u > -22$\,mag. Thus, we only select massive galaxies with a denser CGM at a fixed distance. In addition, these galaxies are assumed to have higher star formation rates and therefore higher RM since star formation can drive magnetized outflows on galactic scales \citep{Wiener_2017, Basu_2018}. The \lss\ sample is more inhomogenous than the COSMOS sample because of the wider redshift range and hence any excess at small impact parameters may be less obvious. The redshift-corrected excess from the combined samples is $3.0\pm 1.2$\,\urm, which is significant at 2.5$\sigma$. 

The increase towards small impact parameters of the |RM| is consistent with previous results suggesting the presence of significant magnetic fields in and around galaxies at distances of several tens of kpc \citep{Bernet_2008, Bernet_2012, Bernet_2013}. However, we find that our results are lower than found in some previous work. \citet{Farnes_2014} found an excess in |RM| of $24\pm 6$\,\urm\ in Mg\,{\sc ii} absorbers and \citet{Bernet_2008} found an excess of 140\,\urm. However, more recently, \citet{Heesen_2023} used LOFAR observations to investigate a sample of nearby galaxies and found an excess in |RM| of $3.7$\,\urm\ with an uncertainty between $\pm 0.9$\,\urm\  and $\pm 1.3$\,\urm\ corresponding to a significance of $2.8\sigma$--$4.1\sigma$ which is in agreement with our results. We note that \citet{Heesen_2023} detected the excess of |RM| only for galaxies along the minor axis of inclined galaxies. For our samples at much higher redshifts, we cannot test this since we have no information about the orientation or inclination.
Our data leads to a significance level of 2.5$\sigma$ for the combination of both fields.

From our data, we can estimate the number of detections that we would need to get to a more reliable 3$\sigma$ significance detection level. As $\sigma$ scales with $\sqrt{n}$ and with given $n_{\mathrm{2.5\sigma}} = 125$ (current combined sample size) we estimate the sample size that we would need for a 3$\sigma$ detection to $n_{\mathrm{3\sigma}}=180$.

Another possibility to increase the significance that would not need as many new detections, would be to build a cleaner sample of hosts. For example, if the background hosts consisted of AGN in clusters then these hosts would have a larger intrinsic scatter in |RM| compared to hosts in poor group or isolated environments. Ideally, the background sources would only consist of AGN with a minimal contribution to the overall |RM| scatter. Alternatively, a weighting scheme could be introduced in order to account for the different underlying populations \citep{Rudnick_2019}.

We do not find a connection between redshift and |RM| which could be due to following reasons: As the \lss\ sample has a much higher redshift range ($0.06 < z < 1.09$) than the COSMOS sample ($0.25 < z < 0.46$), the trend of the \lss\ sample provides an indication over a wider redshift range. However, the explicitly high redshift detections of the \lss\ lead to the fact that more intervening material occurs along the line of sight and the sample becomes more scrambled. In contrast, the smaller redshift range of the COSMOS sample leads to a more homogeneous sample so the results derived from this sample has more significance compared to the more heterogenous \lss\ sample. We note that there are two outliers showing high |RM| and high redshift. To test if the positive correlation is only due to those outliers, we mask these two outliers and compute the correlation again. As a result, we still detect a positive trend but with a shallower slope.

Previous work found contradictory results. Work by \citet{Kronberg_1982, Welter_1984} and \citet{Kronberg_2008} have found an increased |RM| at higher redshifts. More recent work by \citet{Hammond_2012, Bernet_2012} and \citet{Pshirkov_2016} did not find any significant evidence for a correlation between |RM| and redshift. We note that the studies that find a positive correlation between |RM| and redshift have been carried out before the release of the re-analyzed NRAO VLA Sky Survey (NVSS) RM catalog \citep{Taylor_2009, Oppermann_2012}. The more recent work that does not find a correlation between |RM| and redshift used the improved NVSS RM catalog.

In order to determine the magnitude of the magnetic field strength around galaxies, we use \cref{RM_eq}. Assuming an electron density of $n_e \approx 10^{-4}\,\rm cm^{-3}$ and a line-of-sight length of $\approx$ 100\,kpc we estimate a magnetic field strength of $B \approx 0.48\, \mu$G for the redshift corrected and combined sample. We note that this result is only a lower limit because the magnetic field strength may be amplified as magnetic field reversals can lower the RM.

The magnetic field strength in the discs of galaxies is usually found to be of order 10--15\,$\mu$G \citep{Pakmor_2017}. Observations of nearby galaxies show that the magnetic field can be described by $B=B_0\exp(-r/r_0)$ \citep{Beck_2015}. Assuming magnetic fields to go down following the above relation from the galactic disk to CGM, and $B_0\approx 10\, \mu$G and $r_0\approx 10$ kpc we derive $B = 0.45\,  \mu$G at a distance of 100\,kpc from the galaxies center. Simulations suggest similar values for the magnetic field strength around galaxies \citep{Pakmor_2020}. \citet{Heesen_2023} found a magnetic field strength of $\approx$$0.50\,  \mu$G for a sample of nearby galaxies. This agrees  with our estimate which is derived from a sample of galaxies at higher $z$ and with higher SFR.

\subsection{The photometric sample}

In the preceding analysis of the spectroscopic catalog, we assumed that the contribution to the RM is dominated by one massive and bright intervenor with the smallest impact parameter to the LOS to the host. However, our host are  located at high redshifts, resulting in a large number of intervening galaxies. Not all galaxies contribute equally to the observed RM. We expect that those with higher masses or higher star-formation rates tend to exhibit stronger magnetic fields in the CGM. Previous studies that examined the RM around galaxies neglected the effect of multiple intervenors.

Using the photometric catalog from \citep{Hatfield_2022} allows us to assess the effect of multiple intervenors on the RM. In addition, by implementing various weighting schemes, we can investigate if galaxies with higher masses or star-formation rates have a greater influence on the RM. In this work, we introduced a weighting scheme that takes into account the star-formation rate of each galaxy. Our initial hypothesis suggests that the RM generated by intervenors is higher when they exhibit higher star-formation rates and if they are located at smaller impact parameters. While our analysis provides some preliminary evidence to support this assumption, we cannot draw any definitive conclusions due to the limitations of the available data. The spectroscopic sample indicates that an excess of |RM| is detected at smaller impact parameters, which fall within the size range of the circumgalactic medium (CGM) when considering only the most luminous galaxy. However, using the the photometric sample, we observe only a small trend between the average impact parameter and the |RM|. As a result, we infer that the most luminous galaxy is most likely the primary contributor to the overall RM. 

\section{Conclusions}\label{Conclusion}

Magnetic fields around galaxies are important to understand galaxy evolution as they regulate the transport of cosmic rays. Direct observations of the CGM are limited to a few nearby sources as the CGM is very tenuous. Indirect methods such as transverse absorption-line studies are therefore a powerful tool to investigate the physical conditions in the CGM such as the metallicity, temperature and gas density and also the magnetic field strength. 

We used MIGHTEE-POL data to measure the RM around foreground star-forming galaxies to investigate the strength of magnetic fields in the CGM. We used catalogs of star-forming and blue cloud galaxies to measure the rotation measure of MIGHTEE-POL sources as a function of the impact parameter from the intervening galaxy and derived the magnetic field strength of the CGM. Also we investigated a possible connection between the |RM| and the redshift. In addition, we studied the impact of all intervenors along the line of sight using the photometric catalogs from \citep{Hatfield_2022} with respect to the mean impact parameter. To account for the impact of mass and star-formation rate of the individual intervenors we introduced a weighting scheme.

In summary, we can draw the following conclusions:

\begin{enumerate}
    \item For a sample with high star-forming galaxies in the MIGHTEE-POL survey by MeerKAT in the \lss\ and COSMOS fields we find an excess of the RM for impact parameters of less than 133 kpc around bright spectroscopic galaxies with a significance of 2.5$\sigma$. We attribute this excess to coherent magnetic fields in the CGM. The excess RM is in agreement with recent work of \citet{Heesen_2023} for nearby galaxies, but lower than previous work \citep{Bernet_2013, Farnes_2014}. We do not subtract the contribution of the Galactic RM.
       
    \item For a complete sample including galaxies down to magnitudes of $M_g \approx - 13.8$, we do not find any RM excess which suggests that only bright, star-forming galaxies with impact parameters less than 130 kpc significantly contribute to the RM of the background radio source.
        
    \item Making rough assumptions on the electron density in the CGM, we estimate the magnetic field strength to be of the order of $B = 0.5\, \mu$G which is in agreement with observations and simulations.
      
    \item We do not find a correlation between the RM of intervening galaxies and their redshift.
     
    \item Using the photometric catalog we do not find a correlation between the total number of intervenors, N$_{int}$, to the total |RM|. Even introducing a weighting scheme that takes into account the star-formation rate does not lead to an |RM| excess in the innermost bin.
\end{enumerate}

Our results suggest that there is a correlation between the impact parameter and the rotation measure which indicates the presence of intervening magnetic fields in the CGM. We find that the |RM| becomes smaller with higher impact parameters for both our samples. Future studies with larger catalogs of background quasars and more accurate RM measurements enable extensive and more detailed studies of the magnetized CGM.

Below we summarize the main caveats of our work:

\begin{enumerate}
\item Selection function of foreground galaxies: The galaxy catalogs that are used are complete in terms of magnitude but 
not volume-limited. Even though the RM from galaxies at higher $z$ contribute less, there may be still undetected intervening galaxies. Certainly for the \lss\ field the coverage is not uniform across the field.

\item The total RM is an integral along the line of sight and we cannot differentiate between contributions from the host, other intervenors or Galactic contributions. Advanced techniques such as rotations measure synthesis can be useful here.

\end{enumerate}


In our work we make use of early-release data from MIGHTEE-POL which is characterized by a relatively high level of uncertainty. Still we could extract statistically significant signals from this early-release data. MIGHTEE serves as a pilot study for surveys with the forthcoming Square Kilometre Array (SKA). Therefore, our results can soon be tested with much larger data sets. The POlarisation Sky Survey of the Universe's Magnetism (POSSUM) carried out by ASKAP, for example, will map a large area of the sky and it is expected to detect millions of rotations measures. Larger catalogs will yield a more robust detection of magnetic fields in the CGM. Prospects for future work also include the study of the CGM of quiescent galaxies, where synchrotron emission due to star formation activity is absent. 

\begin{acknowledgements}
The authors thank the anonymous referee for useful comments and suggestions.\\
The MeerKAT telescope is operated by the South African Radio Astronomy Observatory, which is a facility of the National Research Foundation, an agency of the Department of Science and Innovation. We acknowledge use of the Inter-University Institute for Data Intensive Astronomy (IDIA) data intensive research cloud for data processing. IDIA is a South African university partnership involving the University of Cape Town, the University of Pretoria and the University of the Western Cape. The authors acknowledge the Centre for High Performance Computing (CHPC), South Africa, for providing computational resources to this research project. \\
KB and MB acknowledge funding by the Deutsche Forschungsgemeinschaft (DFG, German Research Foundation) under Germany’s Excellence Strategy – EXC 2121 „Quantum Universe“ – 390833306. \\
KB and MB thank Francesco Sinigaglia and collaborators for providing the catalog used in Sinigaglia et al. 2022.
MNT acknowledges support from the Oxford Hintze Centre for Astrophysical Surveys which is funded through generous support from the Hintze Family Charitable Foundation.  NJA acknowledges support from the ERC Advanced Investigator Grant EPOCHS (788113). RB acknowledges support from an STFC Ernest Rutherford Fellowship [grant number ST/T003596/1]. SPO acknowledges support from the Comunidad de Madrid Atracción de Talento program via grant 2022-T1/TIC-23797. 
\end{acknowledgements}

%
%

\bibliographystyle{aa}
\bibliography{Bib.bib}


\end{document}